\documentclass[12pt,thmsa,a4paper,notitlepage]{article}
\usepackage{sw20lart}

\input{tcilatex}

\begin{document}

\author{Meriem-Hadjer LAGRAA\thanks{%
e-mail : meriem.lagraa@gmail.com } and Mohammed LAGRAA\thanks{%
e-mail : m.lagraa@lycos.com} \\
\textit{Laboratoire de physique th\'{e}orique d'Oran,}\\
\textit{Universit\'{e} d'Oran ,}\\
\textit{31100 Es-S\'{e}nia. ALGERIA.}}
\title{ON THE GENERALIZED EINSTEIN-CARTAN ACTION WITH FERMIONS }
\maketitle

\begin{abstract}
From the freedom exhibited by the generalized Einstein action proposed in 
\cite{Michel-mohamed}, we show that we can construct the standard effective
Einstein-Cartan action coupled to the fermionic matter without the usual
current-current interaction and therefore an effective action which does not
depend neither on the Immirzi parameter nor on the torsion. This establishes
the equivalence between the Einstein-Cartan theory and the theory of the
general relativity minimally coupled to the fermionic matter.
\end{abstract}

In the last few years, the "tetrad-connection" form of the Einstein
gravitational action modified by the Holst term \cite{Holst} is used as a
starting point of the non-perturbative quantization of gravity \cite%
{Ashtekar} \cite{Rovelli}, \cite{Thiemann} (and references therein). This
modified action, which depends on a new dimensionless parameter known as the
Barbero-Immirzi (BI) parameter \cite{Barbero}, \cite{Immirzi}, does not
modify the classical vacuum Einstein equations. But in presence of minimally
coupling of fermionic matter a nonvanishing torsion emerges leading to
equations of motion which depend on this parameter \cite{Perez-Rovelli}, 
\cite{Freidel}, \cite{Bojowald}. This nonvanishing torsion appears in the
effective action under the form of a current-current interaction term with a
coupling constant determined by the BI parameter. Thus, even at the
classical level the effects of this parameter must be observed. This leads
to bring up the question about its physical origin which is still debated in
many works where different interpretations are proposed.

In vacuum, a possible interpretation of the BI parameter as an analog of the 
$\theta $ parameter that describes the different sectors associated to the
topological structure of large gauge transformation in Yang-Mills theory is
proposed in \cite{Gambini}. In \cite{MercurI1} this analogy is investigated
within the framework of the Einstein-Cartan theory nonminimally coupled to
the fermionic matter where it is shown that when the equation of motion of
the connection is satisfied the sum of the Holst term and the nonminimal
term of the fermionic sector reduces to a term where the BI parameter
becomes a coefficient of the Nieh-Yan \cite{Nieh-Yan} topological invariant
which is a total divergence not affecting the equations of motion and
therefore does not modify the usual second order effective action of
Einstein-Cartan coupled to the fermionic matter. This topological
interpretation of the BI parameter is analyzed in \cite{Mercuri2}, \cite%
{Date} where the topological term is played by the Nieh-Yan density. It is
also shown in \cite{Alexandrov} that the BI parameter is not detectable in
classical theory even after nonminimal coupling of fermions. Much more, the
BI parameter is promoted to be a field interacting with gravity rather than
a constant in \cite{Taveras}, \cite{Torres}, \cite{Galcagni}, \cite{Mercuri4}%
, especially in \cite{Mercuri3} where it is shown that the chiral anomaly is
reabsorbed by a redefinition of the BI field. More recently a different
interpretation is proposed in \cite{Smolin} where, from the extension of the
Plebanski theory, it is shown that the BI parameter is related to the
cosmological constant.

In the above-mentioned works the incorporation of the nonminimal coupling of
the fermionic matter into the Einstein-Cartan theory permits to discard the
effects of the BI parameter but not the current-current interaction term in
the effective action. This makes the physical predictions of the
Einstein-Cartan theory different from those of the theory of general
relativity which does not contain spin-spin interaction induced by a
nonvanishing torsion. Let us recall that, up to now, the experimental
successes of the standard model of particle physics and of the theory of
general relativity seem to be in favor of the minimal coupling procedure
free of the torsion of the connection.

In a previous paper \cite{Michel-mohamed}, the Einstein-Cartan action is
generalized by adding an infinity of non trivial local actions which lead to
the classical vacuum Einstein equations. In this paper we show that we can
deduce, from this generalized action, the usual second order effective
action of Einstein-Cartan theory minimally coupled to fermionic matter
without current-current interaction term nor BI parameter-dependent and
therefore an Einstein-Cartan theory minimally coupled to fermionic matter
equivalent to the theory of general relativity with fermions where the
torsion of the connection is assumed to vanish.

We start with the gravity action proposed in \cite{Michel-mohamed} which we
couple minimally to the standard real Dirac Lagrangian

\begin{eqnarray}
S(e,\omega ,\Psi ) &=&\frac{1}{2k}\dint_{\mathcal{M}}\left( \frac{1}{2}%
\epsilon _{IJKL}e^{I}\wedge e^{J}\wedge \Omega ^{KL}+\frac{1}{\gamma }\Theta
^{I}\wedge \Theta _{I}+F(\Theta ^{2})\epsilon \right)  \nonumber \\
&&-\frac{i}{2}\dint_{\mathcal{M}}\ast e_{I}\wedge \left( \overline{\Psi }%
\gamma ^{I}\mathcal{D}\Psi -c.c\right) .  \label{(lagrtotex)}
\end{eqnarray}%
\qquad

Here $\mathcal{M}$ is the $4$-dimensional space-time manifold, $I,$ $J...\in %
\left[ 0,1,2,3\right] $ are internal Lorentz indices, $\epsilon _{IJKL}$ are
the components of the totally antisymmetric Levi-Cevita symbol, $\epsilon
_{0123}=1=-\epsilon ^{0123}$ and $e^{I}=e_{\mu }^{I}dx^{\mu }$ is the
co-tetrad one-form valued in the vectorial representation space endowed with
the flat metric $\eta _{IJ}=diag(-1,1,1,1)$. The metric $\eta _{IJ}$ and its
inverse $\eta ^{IJ}$ are used to lower and to lift the Lorentz indices and
to determine the metric $g_{\mu \nu }=e_{\mu }^{I}e_{\nu }^{J}\eta _{IJ}$ of
the tangent space of the manifold $\mathcal{M}$. $\Omega ^{IJ}$ $=d\omega
^{IJ}+\omega _{\text{ }N}^{I}\wedge \omega ^{NJ}$ is the curvature two-form
associated to the connection one-form $\omega ^{IJ}=\omega _{\mu
}^{IJ}dx^{\mu }=\omega _{\text{ \ \ }K}^{IJ}\wedge e^{K}$ valued in the
Lorentz Lie algebra and $k=8\pi G$ where $G$ is Newton's gravitational
constant in unit $c=1$.

In the second term of (\ref{(lagrtotex)}) 
\[
\Theta ^{I}=\mathcal{D}e^{I}=de^{I}+\omega _{\text{ }K}^{I}\wedge
e^{K}=\Theta _{\text{ }KL}^{I}e^{K}\wedge e^{L} 
\]%
is the torsion two-form and $\mathcal{D}$ is the exterior covariant
derivative. From the identity

\[
\Theta ^{I}\wedge \Theta _{I}=d\left( e^{I}\wedge \mathcal{D}e_{I}\right)
+e^{I}\wedge \mathcal{DD}e_{I}=d\left( e^{I}\wedge \mathcal{D}e_{I}\right)
+e^{I}\wedge e^{J}\wedge \Omega _{IJ} 
\]%
we see that this term is equivalent, up to the surface term which is the
Nieh-Yan \cite{Nieh-Yan} topological invariant, to the one of Holst. Hence $%
\gamma $ can be identified to the BI parameter assumed to be real in the
following.

In the third term of the action (\ref{(lagrtotex)}), $F$ is a $C^{\infty }$
real function of the scalar $\Theta ^{2}$ given by $\Theta ^{I}\wedge \ast
\Theta _{I}=\Theta ^{2}\epsilon $, where $\varepsilon =\frac{1}{4!}\epsilon
_{IJKL}e^{I}\wedge e^{J}\wedge e^{K}\wedge e^{L}=ed^{4}x=\ast 1$ is the
volume form and $e=\det (e)$ is the determinant of $e_{\mu }^{I}$.

In the tetrad formalism, the dual map $\ast $ acts as

\begin{equation}
\ast \left( e^{I_{1}}\wedge ...\wedge e^{I_{q}}\right) =\frac{1}{\left(
4-q\right) !}\epsilon _{\text{ \ \ \ \ \ \ \ \ \ }%
I_{q+1}...I_{4}}^{I_{1}...I_{q}}e^{I_{q+1}}\wedge ...\wedge e^{I_{4}}
\label{(Dualmap)}
\end{equation}%
which by linearity determines the action $\ast $ on any differential form.

Finally in the fourth term, $\gamma ^{I}$ are Dirac matrices and the
exterior covariant derivative $\mathcal{D}$ acts on the Dirac spinors $\Psi $
as $\mathcal{D}\Psi =d\Psi +\frac{\omega ^{IJ}}{2}\sigma _{IJ}\Psi $ where $%
\sigma _{IJ}=$ $\frac{1}{4}\left[ \gamma _{I},\gamma _{J}\right] $ are the
generators of the Lie algebra of the Lorentz group in the Dirac spinorial
representation. "$c.c$" indicates the complex conjugate of the preceding
term.

Due to the signature of the metric $\eta _{IJ}=\frac{1}{2}\left( \gamma
_{I}\gamma _{J}+\gamma _{J}\gamma _{I}\right) $, the Dirac matrices satisfy
the following properties: $\left( \gamma _{0}\right) ^{2}=-I$, $\left(
\gamma _{i}\right) ^{2}=I$, $\gamma _{I}^{\dag }=\gamma _{0}\gamma
_{I}\gamma _{0}$, $\gamma _{0}\sigma _{KL}^{\dag }=-\sigma _{KL}\gamma _{0}$
leading to $\mathcal{D}\overline{\Psi }=d\overline{\Psi }-\overline{\Psi }%
\sigma _{IJ}\frac{\omega ^{IJ}}{2}$ where $\overline{\Psi }=\Psi ^{\dagger
}\gamma ^{0}$ with $\Psi ^{\dagger }$ is the Hermitian conjugation of the
column $\Psi $. These Dirac matrix properties lead to the relations

\[
\left[ \gamma _{M},\sigma _{KL}\right] _{+}=-i\epsilon _{MKLN}\gamma
_{5}\gamma ^{N}\text{ \ and \ }\left[ \gamma _{M},\sigma _{KL}\right] =\eta
_{MK}\gamma _{L}-\eta _{ML}\gamma _{K} 
\]%
where $\gamma _{5}=i\gamma _{0}\gamma _{1}\gamma _{2}\gamma _{3}$ satisfying 
$\gamma _{5}^{\dag }=\gamma _{5}$ and $\left( \gamma _{5}\right) ^{2}=I$.

With respect to an arbitrary variation of the connection, the principle of
least action gives the following equation

\begin{equation}
\left( \left( \frac{1}{2}\epsilon _{IJ}^{\text{ \ \ }KL}+\delta _{\left[ IJ%
\right] }^{KL}\frac{1}{\gamma }\right) \Theta _{\text{ }NM}^{I}+\frac{1}{2}%
F^{\prime }\delta _{\left[ IJ\right] }^{KL}\Theta _{\text{ }PQ}^{I}\epsilon
_{\text{ \ \ \ }NM}^{PQ}-\frac{k}{4}\delta _{\left[ NM\right] }^{KL}\mathcal{%
J}_{J}\right) e^{N}\wedge e^{M}\wedge e^{J}=0.  \label{freetorsion}
\end{equation}%
where$\ \delta _{\left[ KL\right] }^{IJ}=\frac{1}{2}\left( \delta
_{K}^{I}\delta _{L}^{J}-\delta _{L}^{I}\delta _{K}^{J}\right) =\delta _{KL}^{%
\left[ IJ\right] }$ and $\mathcal{J}^{I}=\overline{\Psi }\gamma _{5}\gamma
^{I}\Psi $ is the axial fermionic current. $F^{\prime }(x)=\frac{dF(x)}{dx}$
denotes the derivative of $F(x)$ where $x=\Theta ^{2}$.

To solve this equation, we decompose the torsion components $\Theta ^{IJK}$
into three disjoint representations of the Lorentz group as

\begin{equation}
\Theta ^{IJK}=\epsilon ^{IJKP}\mathcal{A}_{P}+\left( \eta ^{IJ}\Lambda
^{K}-\eta ^{IK}\Lambda ^{J}\right) +T^{IJK}  \label{(decomptors)}
\end{equation}%
where $\epsilon ^{IJKP}\mathcal{A}_{P}$ is completely antisymmetric in $I$, $%
J$, $K$, $\Lambda ^{I}$ is the trace component and $T^{IJK}$ has vanishing
trace and vanishing completely antisymmetric projection, i.e., $%
T^{IJK}+T^{IKJ}=0$, $\eta _{IJ}T^{IJK}=0$ and $T^{IJK}+T^{JKI}+T^{KIJ}=0$.

From the substitution of (\ref{(decomptors)})\ into the equation (\ref%
{freetorsion}), we obtain

\begin{eqnarray}
&&\epsilon ^{RKLP}\left( \left( 1-2F^{\prime }\right) \mathcal{A}_{P}+\frac{2%
}{\gamma }\Lambda _{P}\right)  \nonumber \\
&&-\left( \eta ^{RK}\left( \left( 2-F^{\prime }\right) \Lambda ^{L}-\frac{1}{%
\gamma }\mathcal{A}^{L}\right) -\eta ^{RL}\left( \left( 2-F^{\prime }\right)
\Lambda ^{K}-\frac{1}{\gamma }\mathcal{A}^{K}\right) \right)  \nonumber \\
&&+T^{INM}\left( \left( 1+F^{\prime }\right) \delta _{I}^{R}\delta
_{N}^{K}\delta _{M}^{L}+\frac{1}{\gamma }\epsilon _{NM}^{\text{ \ \ \ \ \ }%
JR}\delta _{\left[ IJ\right] }^{KL}\right) =-\frac{k}{4}\epsilon ^{RKLP}%
\mathcal{J}_{P}.  \label{(freetorsedec)}
\end{eqnarray}

Contracting this equation with $\epsilon _{RKLJ}$ then with $\eta _{RK}$ we
get

\begin{equation}
\left( \left( 1-2F^{\prime }\right) \mathcal{A}_{J}+\frac{2}{\gamma }\Lambda
_{J}\right) =-\frac{k}{4}\mathcal{J}_{J}\text{ }  \label{(antisymcomp)}
\end{equation}%
and%
\begin{equation}
\text{\ }\left( 2-F^{\prime }\right) \Lambda ^{L}-\frac{1}{\gamma }\mathcal{A%
}^{L}=0  \label{(vectcomp)}
\end{equation}%
respectively. Inserting (\ref{(antisymcomp)}) and (\ref{(vectcomp)}) into (%
\ref{(freetorsedec)}) we get%
\begin{equation}
T^{INM}\left( \left( 1+F^{\prime }\right) \delta _{I}^{R}\delta
_{N}^{K}\delta _{M}^{L}+\frac{1}{\gamma }\epsilon _{NM}^{\text{ \ \ \ \ \ }%
JR}\delta _{\left[ IJ\right] }^{KL}\right) =0.  \label{(tracecomp)}
\end{equation}

By combining (\ref{(antisymcomp)}) with (\ref{(vectcomp)}) we obtain the
irreducible components of the torsion in terms of the axial current as:

\begin{equation}
\Lambda ^{L}=-\frac{k}{8\gamma B}\mathcal{J}^{L},\text{ }\mathcal{A}^{L}=%
\frac{k\left( F^{\prime }-2\right) }{8B}\mathcal{J}^{L}
\label{(solutorsion)}
\end{equation}%
defined when $B=\left( F^{\prime }-\frac{1}{2}\right) \left( F^{\prime
}-2\right) +\frac{1}{\gamma ^{2}}\neq 0$ from which follows that the
function $F$ is such that the image $\func{Im}(F^{\prime })$ of $F^{\prime }$
is disjoint from the roots $r_{1}$and $r_{2}$ of the equation $B=0$.

By contracting (\ref{(tracecomp)}) with $\left( 1+F^{\prime }\right) \delta
_{R}^{S}\delta _{K}^{P}\delta _{L}^{Q}-\frac{1}{\gamma }\epsilon _{KL}^{%
\text{ \ \ \ \ \ }TS}\delta _{\left[ RT\right] }^{PQ}$ we get 
\begin{equation}
T^{IKL}\left( \left( 1+F^{\prime }\right) ^{2}+\frac{1}{\gamma ^{2}}\right)
=0  \label{(Tcomp)}
\end{equation}%
leading to 
\begin{equation}
T^{IJK}=0.  \label{(solutortrace)}
\end{equation}%
\qquad

Recall that the parameter $\gamma $ and the function $F$ are assumed to be
real in this paper.

One must notice, however, that the first equation of (\ref{(solutorsion)})
violates the parity transformation. The trace component $\Lambda ^{L}$ of
the torsion has to be a proper vector but according to this equation it
turns out to be proportional to an axial spinor current. As we show below,
this apparent inconsistency, caused by the Holst term which is not invariant
under the parity transformation, does not affect the effective action. In
fact, in its absence which corresponds to the limiting case $\gamma
\longrightarrow \infty $, we deduce from (\ref{(Tcomp)}), (\ref{(vectcomp)})
and (\ref{(antisymcomp)}) the equations 
\begin{equation}
T^{IJK}=0\text{, }\Lambda ^{L}=0\text{, }\mathcal{A}^{L}=\frac{k}{8\left(
F^{\prime }-\frac{1}{2}\right) }\mathcal{J}^{L}  \label{(torsionA)}
\end{equation}%
which are invariant under the parity transformation and defined when the
image $\func{Im}(F^{\prime })$ of $F^{\prime }$ is disjoint of the points $%
-1 $, $2$ and $\frac{1}{2}$. This condition is the same as the one of the
pure gravitational sector where the Einstein-Cartan theory becomes
equivalent to the theory of general relativity \cite{Michel-mohamed}.

In order to determine the contributions coming from the different
irreducible components (\ref{(decomptors)}) of the torsion to the action (%
\ref{(lagrtotex)}), we have to solve the structure equation $\mathcal{D}%
e^{I}=de^{I}+\omega _{\text{ }J}^{I}\wedge e^{J}=\Theta ^{I}$ by splitting
the connection $\omega ^{IJ}$ into two parts $\omega ^{IJ}=\widetilde{\omega 
}\left( e\right) ^{IJ}+C^{IJ}$ where $\widetilde{\omega }^{IJ}(e)$ is the
uniquely defined torsion-free $so(3,1)$ spin connection compatible with the
tetrad, $\widetilde{\mathcal{D}}e^{I}=de^{I}+\widetilde{\omega }_{\text{ }%
J}^{I}\wedge e^{J}=0$, and $C^{IJ}=C_{\text{ \ }\mu }^{IJ}dx^{\mu }=C_{\text{
\ \ }K}^{IJ}e^{K}$ is the contortion one-form.

To get a metric connection, $\mathcal{D}\eta ^{IJ}=0$, the contortion must
satisfy $C^{IJ}+C^{JI}=0$. It is explicitly given in terms of torsion by

\begin{equation}
C^{IJK}=\left( \Theta ^{JIK}+\Theta ^{KIJ}-\Theta ^{IJK}\right)
\label{(Contorsion)}
\end{equation}%
from which we deduce%
\[
\mathcal{D}e^{I}=\widetilde{\mathcal{D}}e^{I}+C_{\text{ }J}^{I}\wedge
e^{J}=C_{\text{ }J}^{I}\wedge e^{J}=C_{\text{ }JK}^{I}e^{K}\wedge
e^{J}=\Theta _{\text{ }JK}^{I}e^{J}\wedge e^{K}=\Theta ^{I}. 
\]

In terms of the irreducible decomposition (\ref{(decomptors)}) of the
torsion, the contortion (\ref{(Contorsion)}) reads

\begin{equation}
C_{\text{ }KL}^{I}=-\epsilon _{\text{ }KL}^{I\text{ \ \ \ \ }P}\mathcal{A}%
_{P}+2\delta _{L}^{I}\Lambda _{K}-2\eta _{KL}\Lambda ^{I}+2T_{L\text{ \ }K}^{%
\text{ \ \ }I}  \label{decompcontor}
\end{equation}%
and 
\begin{equation}
\Theta ^{I}\wedge \ast \Theta _{I}=\Theta ^{2}\epsilon =\left( -12\mathcal{A}%
^{I}\mathcal{A}_{I}+12\Lambda ^{I}\Lambda _{I}+2T^{IJK}T_{IJK}\right)
\epsilon  \label{(theta2)}
\end{equation}%
where we have used the dual map (\ref{(Dualmap)}) and $e^{I}\wedge
e^{J}\wedge e^{K}\wedge e^{L}=-\epsilon ^{IJKL}\epsilon $.

The relation between the curvature $\Omega ^{IJ}=d\omega ^{IJ}+\omega _{%
\text{ }N}^{I}\wedge \omega ^{NJ}$ associated to the connection $\omega
^{IJ} $, the contortion $C^{IJ}$ and the curvature $\widetilde{\Omega }%
^{IJ}=d\widetilde{\omega }^{IJ}+\widetilde{\omega }_{\text{ }N}^{I}\wedge 
\widetilde{\omega }^{NJ}$ associated to the torsion-free $so(1,3)$ spin
connection $\widetilde{\omega }^{IJ}\left( e\right) $ is obtained by acting
the square of the exterior covariant derivative $\mathcal{D}$ on the
vectorial representation $V^{I}$ as

\begin{equation}
\mathcal{DD}V^{I}=\Omega _{\text{ }J}^{I}V^{J}=\left( \widetilde{\Omega }_{%
\text{ }J}^{I}+\widetilde{\mathcal{D}}C_{\text{ }J}^{I}+C_{\text{ }%
N}^{I}\wedge C_{\text{ }J}^{N}\right) V^{J}.  \label{(DDVI)}
\end{equation}

In terms of the torsion-free $so(1,3)$ spin connection and of the
contortion, the action (\ref{(lagrtotex)}) reads

\begin{eqnarray}
S(e,\omega ,\Psi ) &=&\frac{1}{2k}\dint_{\mathcal{M}}\frac{1}{2}\epsilon
_{IJKL}e^{I}\wedge e^{J}\wedge \widetilde{\Omega }^{KL}-\frac{i}{2}\dint_{%
\mathcal{M}}\ast e_{I}\wedge \left( \overline{\Psi }\gamma ^{I}\widetilde{%
\mathcal{D}}\Psi -c.c\right)  \nonumber \\
&&+\frac{1}{2k}\dint_{\mathcal{M}}\left( \frac{1}{2}\epsilon
_{IJKL}e^{I}\wedge e^{J}\wedge C_{\text{ }N}^{K}\wedge C^{NL}+\frac{1}{%
\gamma }\Theta ^{I}\wedge \Theta _{I}+F(\Theta ^{2})\epsilon \right) 
\nonumber \\
&&+\frac{1}{4}\dint_{\mathcal{M}}e^{I}\wedge e^{J}\wedge e^{K}\wedge C_{IJ}%
\mathcal{J}_{K}  \label{decompaction}
\end{eqnarray}%
where the contribution of the term $\widetilde{\mathcal{D}}C_{\text{ }}^{IJ}$
of (\ref{(DDVI)}) is ignored since it reduces to a total derivative due to
the fact that $\widetilde{\mathcal{D}}e^{I}=0$. In (\ref{decompaction}), the
first line of the action describes the torsion-free part. The second and the
third line describe the coupling of the axial fermionic current with the
non-propagating torsion.

The substitution of (\ref{(decomptors)}) and (\ref{decompcontor}) into (\ref%
{decompaction}) gives the explicit form of the contributions coming from the
irreducible components of the torsion

\begin{eqnarray}
S(e,\omega ,\Psi ) &=&\frac{1}{2k}\dint_{\mathcal{M}}\frac{1}{2}\epsilon
_{IJKL}e^{I}\wedge e^{J}\wedge \widetilde{\Omega }^{KL}-\frac{i}{2}\dint_{%
\mathcal{M}}\ast e_{I}\wedge \left( \overline{\Psi }\gamma ^{I}\widetilde{%
\mathcal{D}}\Psi -c.c\right)  \nonumber \\
&&+\frac{1}{2k}\dint_{\mathcal{M}}\frac{1}{\gamma }\left( 24\mathcal{A}%
^{I}\Lambda _{I}-\epsilon ^{IJKL}T_{\text{ \ }IJ}^{N}T_{NKL}\right) \epsilon
\nonumber \\
&&+\frac{1}{2k}\dint_{\mathcal{M}}F(-12\mathcal{A}^{I}\mathcal{A}%
_{I}+12\Lambda ^{I}\Lambda _{I}+2T^{IJK}T_{IJK})\epsilon  \nonumber \\
&&+\frac{1}{2k}\dint_{\mathcal{M}}\left( -24\Lambda ^{I}\Lambda _{I}+6%
\mathcal{A}^{I}\mathcal{A}_{I}-4T^{IJK}T_{KIJ}\right) \epsilon  \nonumber \\
&&+\frac{3!}{4}\dint_{\mathcal{M}}\mathcal{A}^{I}\mathcal{J}_{I}\epsilon .
\label{(Action-Tors)}
\end{eqnarray}

It is easy to see that an arbitrary variation of (\ref{(Action-Tors)}) with
respect to $\mathcal{A}_{I}$, $\Lambda _{I}$ and $T_{IJK}$ gives the
expressions of the irreducible components of the torsion in terms of the
axial current (\ref{(solutorsion)}) and (\ref{(solutortrace)}).

Now, we are ready to discuss the action (\ref{(Action-Tors)}). We start by
verifying the usual case $F=0$ which, from (\ref{(solutorsion)}), gives%
\begin{equation}
\text{ }\Lambda ^{L}=-\frac{k\gamma }{8\left( 1+\gamma ^{2}\right) }\mathcal{%
J}^{L},\text{ }\mathcal{A}^{L}=-\frac{2k\gamma ^{2}}{8\left( 1+\gamma
^{2}\right) }\mathcal{J}^{L}.  \label{(TorsionR)}
\end{equation}
\qquad

By substituting these values of torsion components and (\ref{(solutortrace)}%
) into (\ref{(Action-Tors)}), we recover the second-order tetrad action of
general relativity action coupled to the fermionic matter \cite{Bojowald}%
\begin{eqnarray*}
S(e,\omega ,\Psi ) &=&\frac{1}{2k}\dint_{\mathcal{M}}\frac{1}{2}\epsilon
_{IJKL}e^{I}\wedge e^{J}\wedge \widetilde{\Omega }^{KL}-\frac{i}{2}\dint_{%
\mathcal{M}}\ast e_{I}\wedge \left( \overline{\Psi }\gamma ^{I}\widetilde{%
\mathcal{D}}\Psi -c.c\right) \\
&&-\frac{3k}{16}\frac{\gamma ^{2}}{\left( 1+\gamma ^{2}\right) }\dint_{%
\mathcal{M}}\mathcal{J}^{I}\mathcal{J}_{I}\epsilon
\end{eqnarray*}%
which exhibits the BI parameter dependence in front of the current-current
coupling.

Note that, although the first equation of (\ref{(TorsionR)}) violates the
parity transformation, it leads to an invariant current-current interaction
term in the effective action.

In the absence of the Holst term, we can consider any dimension $d$ of the
space time manifold $\mathcal{M}$ where the action (\ref{(lagrtotex)}) and (%
\ref{(Action-Tors)}) read respectively%
\begin{equation}
S(e,\omega ,\Psi )=\frac{1}{2k}\dint_{\mathcal{M}}\left( \ast \left(
e_{K}\wedge e_{L}\right) \wedge \Omega ^{KL}+F(\Theta ^{2})\epsilon \right) -%
\frac{i}{2}\dint_{\mathcal{M}}\ast e_{I}\wedge \left( \overline{\Psi }\gamma
^{I}\mathcal{D}\Psi -c.c\right)  \label{(lagrtotshos)}
\end{equation}%
and 
\begin{eqnarray}
S(e,\omega ,\Psi ) &=&\frac{1}{2k}\dint_{\mathcal{M}}\ast \left( e_{K}\wedge
e_{L}\right) \wedge \widetilde{\Omega }^{KL}-\frac{i}{2}\dint_{\mathcal{M}%
}\ast e_{I}\wedge \left( \overline{\Psi }\gamma ^{I}\widetilde{\mathcal{D}}%
\Psi -c.c\right)  \nonumber \\
&&+\frac{1}{2k}\dint_{\mathcal{M}}F(2\mathcal{A}^{IJK}\mathcal{A}%
_{IJK}+4\left( d-1\right) \Lambda ^{I}\Lambda _{I}+2T^{IJK}T_{IJK})\epsilon 
\nonumber \\
&&+\frac{1}{2k}\dint_{\mathcal{M}}\left( -4\left( d-1\right) \left(
d-2\right) \Lambda ^{I}\Lambda _{I}-\mathcal{A}^{IJK}\mathcal{A}%
_{IJK}-4T^{IJK}T_{KIJ}\right) \epsilon  \nonumber \\
&&+\frac{\left( -\right) ^{d-1}}{4}\dint_{\mathcal{M}}\mathcal{A}^{IJK}%
\mathcal{J}_{IJK}\epsilon  \label{(ActionDIM)}
\end{eqnarray}%
where $\mathcal{A}^{IJK}$ is the completely antisymmetric component of the
irreducible decomposition (\ref{(decomptors)}) of the torsion and $\mathcal{J%
}^{IJK}=i\overline{\Psi }\left[ \gamma ^{I},\sigma ^{JK}\right] _{+}\Psi $
is the tensorial current which is also completely antisymmetric in $I$, $J$, 
$K$.

An arbitrary variation of (\ref{(ActionDIM)}) with respect to $T^{IJK}$, $%
\Lambda ^{I}$and $\mathcal{A}^{IJK}$ gives

\begin{equation}
T^{IJK}=0\text{, }\Lambda ^{I}=0\text{, }\mathcal{A}^{IJK}=\frac{\left(
-\right) ^{d}k}{8\left( F^{\prime }-\frac{1}{2}\right) }\mathcal{J}^{IJK}
\label{(soltortiondim)}
\end{equation}%
defined if the function $F$ is such that the image $\func{Im}(F^{\prime })$
of $F^{\prime }$ is disjoint of the points $-1$, $d-2$ and $\frac{1}{2}$. By
inserting these values of the torsion components into (\ref{(ActionDIM)}),
we get

\[
S(e,\omega ,\Psi )=\frac{1}{2k}\dint_{\mathcal{M}}\ast \left( e_{K}\wedge
e_{L}\right) \wedge \widetilde{\Omega }^{KL}-\frac{i}{2}\dint_{\mathcal{M}%
}\ast e_{I}\wedge \left( \overline{\Psi }\gamma ^{I}\widetilde{\mathcal{D}}%
\Psi -c.c\right) 
\]%
\[
+\frac{1}{2k}\dint_{\mathcal{M}}\left( F(2\mathcal{A}^{IJK}\mathcal{A}%
_{IJK})-4\mathcal{A}^{IJK}\mathcal{A}_{IJK}F^{\prime }\left( 2\mathcal{A}%
^{IJK}\mathcal{A}_{IJK}\right) +\mathcal{A}^{IJK}\mathcal{A}_{IJK}\right)
\epsilon 
\]%
where we have replaced in the last term of the action (\ref{(ActionDIM)})
the tensorial current by its expression in term of $\mathcal{A}^{IJK}$, $%
\mathcal{J}^{IJK}=\left( -\right) ^{d}\frac{8}{k}\left( F^{\prime }-\frac{1}{%
2}\right) \mathcal{A}^{IJK}$. If we try to get an action without
current-current interaction, and then an equivalence between the
Einstein-Cartan theory and the theory of general relativity, we have to
cancel the second line of this action which consists to solve the equation

\begin{equation}
F(x)-2xF^{\prime }\left( x\right) +\frac{x}{2}=0  \label{(equationtors)}
\end{equation}%
where $x=2\mathcal{A}^{IJK}\mathcal{A}_{IJK}$. The smooth solution $F(x)=%
\frac{x}{2}$ of this equation is not acceptable because it contradicts the
condition $F^{\prime }-\frac{1}{2}\neq 0$ required to have a well defined
relation between the torsion and the fermionic current given in the equation
(\ref{(soltortiondim)}) but if we accept real nonpolynomial solutions, the
general solution\footnote{%
We thank the referee for suggesting this solution} of (\ref{(equationtors)})
is $F(x)=\frac{x}{2}+c\sqrt{x}$ where $c$ is any real scalar function which
does not depend on the torsion. This solution is defined when $x=2\mathcal{A}%
^{IJK}\mathcal{A}_{IJK}\rangle 0$.

This case, $F^{\prime }-\frac{1}{2}=\frac{c}{2\sqrt{2\mathcal{A}^{IJK}%
\mathcal{A}_{IJK}}}\neq 0$, leads to a no singular value of the completely
antisymmetric irreducible component of the torsion 
\[
\frac{\left( -\right) ^{d}4c}{k\sqrt{2\mathcal{A}^{NML}\mathcal{A}_{NML}}}%
\mathcal{A}^{IJK}=\mathcal{J}^{IJK} 
\]%
which reduces, for $d=4$, to%
\[
\frac{4c}{k\sqrt{-12\mathcal{A}^{K}\mathcal{A}_{K}}}\mathcal{A}^{I}=\mathcal{%
J}^{I} 
\]%
defined if the vector $\mathcal{A}^{L}$ is time-like ,i.e. $\mathcal{A}^{I}%
\mathcal{A}_{I}\langle 0$.

Therefore, with this nonpolynomial choice of the function $F$, the action (%
\ref{(lagrtotshos)}) reads%
\begin{eqnarray*}
S(e,\omega ,\Psi ) &=&\frac{1}{2k}\dint_{\mathcal{M}}\left( \ast \left(
e_{K}\wedge e_{L}\right) \wedge \Omega ^{KL}+\left( \frac{\Theta ^{2}}{2}+c%
\sqrt{\Theta ^{2}}\right) \epsilon \right) \\
&&-\frac{i}{2}\dint_{\mathcal{M}}\ast e_{I}\wedge \left( \overline{\Psi }%
\gamma ^{I}\mathcal{D}\Psi -c.c\right)
\end{eqnarray*}
leading to the effective action

\[
S_{eff}(e,\Psi )=\frac{1}{2k}\dint_{\mathcal{M}}\left( \ast \left(
e_{K}\wedge e_{L}\right) \wedge \widetilde{\Omega }^{KL}\right) -\frac{i}{2}%
\dint_{\mathcal{M}}\ast e_{I}\wedge \left( \overline{\Psi }\gamma ^{I}%
\widetilde{\mathcal{D}}\Psi -c.c\right) 
\]%
which is the standard second-order tetrad form of the Einstein-Cartan theory
minimally coupled to the fermionic matter without the current-current term
and which does not depend neither on the torsion nor on the arbitrary
function $c$.

Now, if we consider the Holst term which work in 4-dimensional space-time
manifold $\mathcal{M}$ only, the relations (\ref{(solutorsion)}) show that
for the regular value $F^{\prime }=2$, $B=\frac{1}{\gamma ^{2}}\neq 0$, we
get 
\begin{equation}
\mathcal{A}^{L}=0,\text{ }T^{IJK}=0,\text{ }\Lambda ^{L}=-\frac{k\gamma }{8}%
\mathcal{J}^{L}  \label{(TvecJs)}
\end{equation}%
which make the action (\ref{(Action-Tors)}) more simple. In fact, from the
solution $F(\Theta ^{2})=2\Theta ^{2}+C^{t}=24\Lambda ^{I}\Lambda _{I}+C^{t}$
of the equation $F^{\prime }=2$, the $F$-term of the action (\ref%
{(lagrtotex)}) writes $F(\Theta ^{2})\epsilon =2\Theta ^{I}\wedge \ast
\Theta _{I}+C^{t}\epsilon $ where the constant is proportional to the
cosmological constant. For gravity without cosmological constant , $C^{t}=0$%
, the action (\ref{(lagrtotex)}) reads

\begin{eqnarray}
S(e,\omega ,\Psi ) &=&\frac{1}{2k}\dint_{\mathcal{M}}\left( \frac{1}{2}%
\epsilon _{IJKL}e^{I}\wedge e^{J}\wedge \Omega ^{KL}+\frac{1}{\gamma }\Theta
^{I}\wedge \Theta _{I}+2\Theta ^{I}\wedge \ast \Theta _{I}\right)  \nonumber
\\
&&-\frac{i}{2}\dint_{\mathcal{M}}\ast e_{I}\wedge \left( \overline{\Psi }%
\gamma ^{I}\mathcal{D}\Psi -c.c\right) .  \label{(lagrtotfinal)}
\end{eqnarray}

In spite of the fact that the nonvanishing torsion is given by the relation (%
\ref{(TvecJs)}) which violates the parity transformation, it does not appear
in the action (\ref{(Action-Tors)}). In fact by inserting $\mathcal{A}^{L}=0$%
, $T^{IJK}=0$ and $F(\Theta ^{2})=2\Theta ^{2}=24\Lambda ^{I}\Lambda _{I}$
into (\ref{(lagrtotfinal)}), which consist to make this insertion into (\ref%
{(Action-Tors)}), we see that the Holst term and the last one which
describes the coupling of the torsion with the current vanish while the
third and the fourth line cancel each other out to reduce to the effective
action%
\begin{equation}
S_{eff}(e,\Psi )=\frac{1}{4k}\dint_{\mathcal{M}}\epsilon _{IJKL}e^{I}\wedge
e^{J}\wedge \widetilde{\Omega }^{KL}-\frac{i}{2}\dint_{\mathcal{M}}\ast
e_{I}\wedge \left( \overline{\Psi }\gamma ^{I}\widetilde{\mathcal{D}}\Psi
-c.c\right)   \label{effectiveaction}
\end{equation}%
which is the standard second-order tetrad form of the Einstein-Cartan theory
minimally coupled to the fermionic matter without the current-current term
and which does not depend neither on the torsion nor on the BI parameter. \ 

Note that although the effective action (\ref{effectiveaction}) does not
depend of the BI parameter, the Holst term is necessary to get a well
defined relation between the trace component of the torsion and the
fermionic current (\ref{(TvecJs)}). In its absence, $\gamma \longrightarrow
\infty $, $F^{\prime }=2$ is a singular point which makes the trace
component undetermined with or without fermionic matter. This undetermined
trace component does not appear to have any physical significance as we see
it from (\ref{(Action-Tors)}); in fact if $\gamma \longrightarrow \infty $
and if we consider the solution $F(\Theta ^{2})=2\Theta ^{2}$ of the
equation $F^{\prime }=2$ the trace component disappears but an arbitrary
variation with respect to $T^{IJK}$ and $\mathcal{A}^{I}$ gives $T^{IJK}=0$
and $\mathcal{A}^{I}=\frac{k}{12}\mathcal{J}^{I}$ leading to an
current-current interaction term $\frac{k}{16}\mathcal{J}^{I}\mathcal{J}_{I}$
in the effective action.

We conclude this work by emphasizing that as opposite to the usual tetrad
form of the gravitational action with fermions where a torsion emerges under
the form of current-current interaction which is either $\gamma $-dependent
in the Holst-modified Einstein-Cartan theory or $\gamma $-independent in
presence of nonminimally coupled fermionic matter, we have shown that if, in
addition to the Holst-modified Einstein-cartan theory, we add a torsion term
of the form $\frac{1}{k}\dint_{\mathcal{M}}\Theta ^{I}\wedge \ast \Theta
_{I} $ (\ref{(lagrtotfinal)}) we obtain an effective Einstein-Cartan theory
minimally coupled to the fermionic matter (\ref{effectiveaction}) where
neither the torsion, which induces the current-current interaction, nor the
BI parameter take place. Therefore, the generalized Einstein-Cartan action (%
\ref{(lagrtotfinal)}) is equivalent to the theory of the general relativity
minimally coupled to the fermionic matter. Note that this equivalence is not
based on a completely vanishing torsion, its irreducible trace component
does not vanish (\ref{(TvecJs)}). This leads to a space-time manifold with a
nonvanishing torsion depending on the BI parameter which does not induce
physical effects because of its absence in the effective action (\ref%
{effectiveaction}).

We have also shown that if we add to the Einstein-Cartan action a
nonpolynomial term in torsion of the form $\frac{1}{2k}\dint_{\mathcal{M}%
}\left( \frac{\Theta ^{2}}{2}+c\sqrt{\Theta ^{2}}\right) \epsilon $ we
obtain an effective action equivalent to the theory of the general
relativity minimally coupled to the fermionic matter. This equivalence is
not based on a vanishing torsion also, its irreducible completely
antisymmetric component does not vanish (\ref{(soltortiondim)}). In spite of
the fact that it is related to the fermionic current, it is an inert
quantity which does not affect the evolution equations. Note that we can see
from (\ref{(soltortiondim)}) that the presence of the fermionic matter is
necessary in this case because in its absence we can not obtain a vanishing
torsion which is a condition to have an equivalence in pure gravitational
sector.

\end{document}